\documentclass{bmcart}

\usepackage[utf8]{inputenc} 

\usepackage{amsmath,verbatim,color,amssymb,upgreek,epsfig,bm,mathrsfs,amsthm,latexsym,subfigure,graphicx}
\usepackage{multirow}
\usepackage[figuresright]{rotating}
\usepackage{array}
\usepackage{colortbl}
\usepackage{soul}
\usepackage{xcolor}

\def\includegraphics{}

\startlocaldefs
\endlocaldefs

\begin{document}

\newcommand{\bcU}{\boldsymbol{\cal U}}
\newcommand{\bbeta}{\boldsymbol{\beta}}
\newcommand{\bdelta}{\boldsymbol{\delta}}
\newcommand{\bDelta}{\boldsymbol{\Delta}}
\newcommand{\boldeta}{\boldsymbol{\eta}}
\newcommand{\bxi}{\boldsymbol{\xi}}
\newcommand{\bfeta}{\boldsymbol{\eta}}
\newcommand{\bGamma}{\boldsymbol{\Gamma}}
\newcommand{\bSigma}{\boldsymbol{\Sigma}}
\newcommand{\balpha}{\boldsymbol{\alpha}}
\newcommand{\bOmega}{\boldsymbol{\Omega}}
\newcommand{\btheta}{\boldsymbol{\theta}}
\newcommand{\bmu}{\boldsymbol{\mu}}
\newcommand{\bnu}{\boldsymbol{\nu}}
\newcommand{\bgamma}{\boldsymbol{\gamma}}
\newcommand{\bpsi}{\boldsymbol{\psi}}
\newcommand{\bphi}{\boldsymbol{\phi}}
\newcommand{\bomega}{\boldsymbol{\omega}}

\newcommand{\redcolor}[1]{{\leavevmode\color{red}#1}}

\renewenvironment{enumerate}
{\begin{list}{\arabic{enumi}.}
      {\setlength{\leftmargin}{2.5em}
       \setlength{\itemsep}{-\parsep}
       \setlength{\topsep}{-\parskip}
       \usecounter{enumi}}
 }{\end{list}}

\newcommand{\comm}[1]{}

\begin{frontmatter}

\begin{fmbox}
\dochead{Research}


\title{Overlapping-sample Mendelian randomisation with multiple exposures: A Bayesian approach}

\author[
   addressref={aff1},
]{\inits{L}\fnm{Linyi} \snm{Zou}}
\author[
   addressref={aff1},                   
   corref={aff1},                       
   email={hui.guo@manchester.ac.uk}   
]{\inits{H}\fnm{Hui} \snm{Guo}}
\author[
   addressref={aff1},
]{\inits{C}\fnm{Carlo} \snm{Berzuini}}


\address[id=aff1]{
  \orgname{Centre for Biostatistics, School of Health Sciences, The University of Manchester}, 
  \street{Oxford Road},                     %
  \postcode{M13 9PL}                                
  \city{Manchester},                              
  \cny{UK}                                    
}


\begin{artnotes}

\end{artnotes}

\end{fmbox}


\begin{abstractbox}

\begin{abstract} 
\parttitle{Background} 
Mendelian randomization (MR) has been widely applied to causal inference in medical research. It uses genetic variants as instrumental variables (IVs) to investigate putative causal relationship between an exposure and an outcome. Traditional MR methods have mainly focussed on a two-sample setting in which IV-exposure association study and IV-outcome association study are independent.
However, it is not uncommon that participants from the two studies fully overlap (one-sample) or partly overlap (overlapping-sample).

\parttitle{Methods} 
We proposed a Bayesian method that is applicable to all the three sample settings. In essence, we converted a two- or overlapping- sample MR to a one-sample MR where data were partly unmeasured. Assume that all study individuals were drawn from the same population and unmeasured data were missing at random. Then the missing data were treated au pair with the model parameters as unknown quantities, and thus, were imputed iteratively conditioning on the observed data and estimated parameters using Markov chain Monte Carlo. We generalised our model to allow for pleiotropy and multiple exposures and assessed its performance by a number of simulations using four metrics: mean, standard deviation, coverage and power. We also compared our method with classic MR methods.

\parttitle{Results} 
In our proposed method, higher sample overlapping rate and instrument strength led to more precise estimated causal effects with higher power. Pleiotropy had a notably negative impact on the estimates. Nevertheless, the coverages were high and our model performed well in all the sample settings overall. In comparison with classic MR, our method provided estimates with higher precision. When the true causal effects were non-zero, power of their estimates was consistently higher from our method. The performance of our method was similar to classic MR in terms of coverage.

\parttitle{Conclusions} 
Our model offers the flexibility of being applicable to any of the sample settings. It is an important addition to the MR literature which has restricted to one- or two- sample scenarios. Given the nature of Bayesian inference, it can be easily extended to more complex MR analysis in medical research.

\end{abstract}




\end{abstractbox}
%

\end{frontmatter}





\section*{Background}

Many questions in medical research address putative causal nature of the relationship between a clinical outcome and a corresponding risk factor or exposure. Randomised controlled trials (RCTs) are ideal for this purpose, but are often infeasible due to cost or ethical considerations. With the aid of proper statistical methodology, causal inference can be made from observational studies.  This is a context where Mendelian randomization (MR) \cite{Martjin1986, George2003,Debbie2008} can play an important role. MR uses genetic variants as instrumental variables (IVs) to estimate the causal effect of an exposure on an outcome of interest \cite{Philip2017, Carlo2018}.
Without loss of generality, we assume the IVs are single nucleotide polymorphisms (SNPs).

MR analysis requires data from two association studies: IV-exposure and IV-outcome. Most of MR methods \cite{Toby2013, Jack2015, Jack2016, Zhao2018} have focussed on a two-sample scenario where the IV-exposure and IV-outcome associations were estimated separately from independent studies. In other words, there were no shared individuals between these two studies. A limited number of one-sample MR methods, if a single group of individuals come with a complete set of measurements (IVs, exposure and outcome), are also available in the literature  \cite{Carlo2018, Stephen2014, Frank2003, Elinor2012}. Little attention has, however, been devoted to overlapping samples where the two studies have a subset of individuals in common. For example, an investigator has identified a number of SNPs associated with diabetes from a genome-wide association study. Her interest is in whether expression levels of certain genes are causal risk factors of diabetes. But gene expression can be measured only from a subset of participants of the SNP-diabetes association study due to cost. Another example is that the investigator has collected data from two small independent IV-BMI and IV-diabetes association studies, with the aim to estimate the causal effect of BMI on diabetes using MR. To enhance statistical power, a natural way is to incorporate previous studies that have complete sets of observed data on IV, BMI and diabetes into her current studies.

To the best of our knowledge, few studies have investigated overlapping-sample MR. Burgess et al. \cite{Burgess2016} have shown that sample overlap increases type I error and leads to bias in classic MR methods. LeBlanc et al. \cite{LeBlanc2018} have developed a correction method of overlapping samples by decorrelation. There is a pressing need for the development of a flexible MR approach that can be used for one-,  two- as well as overlapping- samples.
Here we propose a novel MR method based on the
Bayesian framework introduced by Berzuini et al \cite{Carlo2018}. Our method preserves data information while
avoiding bias. It exploits two simple, but powerful, ideas:
 (i) the overlapping-sample problem is a special case
of the more general situation where some or all
the individuals have missing
values in exposure or in outcome;
and (ii)
a Bayesian approach can coherently deal with missing
data by treating them as additional parameters which can be estimated
from the data. In this paper we introduce our method and
assess its performance by comparing it with classic MR in a simulation study. A further
advantage of Bayesian MR
is freedom to elaborate the model
in various directions of interest.
This feature we illustrate by moving away from the standard single-exposure MR model and considering multiple exposures instead.

\section*{Methods}
\subsection*{Model}
Let $U$ denote a set of unobserved confounders which could possibly distort the causal relationship between the exposure $X$ and the outcome $Y$. Let $Z$ be an instrumental SNP (multiple instruments are considered later) which satisfies the following three assumptions (\cite{Stephen2014}):

\begin{itemize}
  \item[] $\rm A_{1}$: $Z$ is associated with $X$;
  \item[] $\rm A_{2}$: $Z$ is independent of $U$;
  \item[] $\rm A_{3}$: $Z$ is independent of $Y$, conditioning on $(X, U)$.
\end{itemize}

These assumptions are graphically expressed in Figure \ref{figure1}.  The $Z \rightarrow X$ arrow represents a non-zero association between the IV and the exposure, in accord with $\rm A_{1}$. Assumption $\rm A_{2}$ follows from the graph (it would not if there were an arrow directly connects $Z$ and $U$). $\rm A_{3}$ is also satisfied (it would not if an arrow pointed directly from $Z$ to $Y$). The $X \rightarrow Y$ arrow represents the putative causal effect of $X$ on $Y$, which is our primary interest.

Consider a study with multiple exposures and each instrumental SNP associated with at least one exposure. Without loss of generality, we shall hereafter restrict attention to the case where there are three exposures $\mathbf{X} = (X_{1}, X_{2}, X_{3})$ and three IVs $\mathbf{Z} = (Z_{1}, Z_{2}, Z_{3})$. Figure \ref{figure2} depicts our model and  parameter notations, with $\beta_{1}$, $\beta_{2}$ and $\beta_{3}$ representing the putative causal effects of the exposures $X_{1}$, $X_{2}$ and $X_{3}$ on the outcome $Y$, respectively; $\delta_{1}$, $\delta_{2}$, $\delta_{3}$ and $\delta_{4}$ representing the effects of the confounder $U$ on the three exposures and $Y$, respectively. For simplicity, we assume that there are no direct associations between the exposures and that the IVs are mutually independent. The $Z_1 \rightarrow X_1$ association is quantified by parameter $\alpha_{1}$ and the $Z_2 \rightarrow X_2$ association by $\alpha_{2}$. Both $Z_{2}$ and $Z_{3}$ are associated with $X_{3}$, and the strengths of these associations are quantified by parameters $\alpha_{3}$ and $\alpha_{4}$ respectively. $Z_{1}$ and $Z_{3}$ are ``valid instruments" as they satisfy all the above stated MR assumptions. However, $Z_{2}$ violates assumption $\rm A_{3}$ because the effect of $Z_{2}$ on $Y$ is mediated by both $X_{2}$ and $X_{3}$ - a problem of horizontal pleiotropy \cite{Jordan2019}.

By assuming linearity and additivity of the conditional dependencies, and
in accord with the graph of Figure 1, we specify our model as follows.
\begin{eqnarray}\label{3.1}
  U &\sim& N(0, 0.1),\\
\label{3.2}
  X_{1}  | Z_{1},U &\sim& N(\omega_{1}+\alpha_{1}Z_{1}+\delta_{1}U, \sigma_{1}^{2}),\\
\label{3.3}
  X_{2} | Z_{2},U &\sim& N(\omega_{2}+\alpha_{2}Z_{2}+\delta_{2}U, \sigma_{2}^{2}),\\
\label{3.4}
  X_{3} | Z_{2},Z_{3},U &\sim& N(\omega_{3}+\alpha_{3}Z_{2}
+\alpha_{4}Z_{3}+\delta_{3}U, \sigma_{3}^{2}),\\
\label{3.5}
  Y | X_{1},X_{2},X_{3},U &\sim& N(\omega_{Y}+\beta_{1}X_{1}+
\beta_{2}X_{2}+\beta_{3}X_{3}+\delta_{4}U, \sigma_{Y}^{2}),
\end{eqnarray}

where $N(a,b)$ stands for a normal distribution with mean $a$ and variance $b$. The $\omega$ parameters are unknown intercepts and the $\sigma$s are standard deviations of independent random noise terms. The parameters of primary inferential interest are the $\beta$s, each representing the causal effect of a particular exposure on the outcome. The $\alpha$ parameters quantify the strengths of pairwise associations between the instruments and the exposures. They are often referred to as ``instrument strengths", and should be significantly different from zero to attenuate  weak instrument bias \cite{Burgess2011}. Finally, $U$ denotes a sufficient scalar summary of the unobserved confounders, which is set to be drawn from a $N(0,0.1)$ distribution with its parameters not identifiable from the likelihood. This model is built on the basis of the Bayesian approach developed by Berzuini et al.  \cite{Carlo2018}, with an extension to multiple exposures and allowing for an IV to be associated with more than one exposure.

\subsection*{Our method}

We propose a Bayesian MR method that works in all the sample settings (whether one-sample, two-sample or overlapping-sample), and therefore is more flexible than existing non-Bayesian methods in this respect.
In essence, our method treats two- or overlapping- samples as one sample where the data of some (or all) of the individuals are incomplete, in the sense that these individuals have $\mathbf{Z}$ and $\mathbf{X}$ (but not $Y$) measured or $\mathbf{Z}$ and $Y$ (but not $\mathbf{X}$) measured. Clearly our method
allows for quite general patterns of missingness, although
it is not our purpose here to explicitly
explore their full repertoire. In addition, for simplicity, we
illustrate our method by restricting attention to the
special case of a continuous outcome variable $Y$.
We assume that all individuals are drawn from the same population and unmeasured data are missing at random. Then whatever are the unobserved variables $\mathbf{X}$ or $Y$, they are treated au pair with the model parameters as unknown quantities, and thus, can be imputed iteratively from their distributions conditioning on the observed variables and estimated parameters using Markov chain Monte Carlo (MCMC) \cite{Nicholas1953} - the best Bayesian tradition.

We start with introducing three disjoint datasets as follows (see Figure \ref{figure3} (a)):
\begin{itemize}
  \item Dataset $A$: all individuals
 with observed values of $\mathbf{Z}$, $\mathbf{X}$ and $Y$;
  \item Dataset $B$:  all  individuals
 with observed values of $\mathbf{Z}$ and $\mathbf{X}$ only;
  \item Dataset $C$: all  individuals
 with observed values of $\mathbf{Z}$ and $Y$ only.
\end{itemize}

No individuals are common in $A$, $B$ and $C$ but they are drawn independently from the same population.  As discussed earlier, $A$ may be collected from a previous study; $B$ and $C$ may represent data from two independent IV-exposure and IV-outcome association studies. If we combine $A$ with $B$
$$  D_{1} = A \cup B$$
and $A$ with $C$
$$  D_{2} = A \cup C,$$
the two resulting datasets,  $  D_{1}$ and $  D_{2}$, will not be disjoint as they have dataset $A$ in common- a typical overlapping-sample problem. If  $A$ is empty,  $B$ and $C$ will form a two-sample problem.

Our method treats missing data as all other unknown quantities in the model. That is, we impute the missing values of $Y$ in $B$ by randomly drawing a value from  the conditional distribution of $Y$ given observed exposures and the estimated parameters at each iteration. Likewise, the missing values of $\mathbf{X}$ in $C$ will be imputed by a randomly draw from the conditional distribution of $\mathbf{X}$ given the observed instruments $\mathbf{Z}$ and the estimated parameters at each iteration. Let $Y^{*}$ be imputed values of $Y$ and $\mathbf{X}^{*} = (X_{1}^{*}$, $X_{2}^{*}$, $X_{3}^{*})$ be imputed values of $\mathbf{X}$.  Given all the datasets $A, B$ and $C$, we proceed with the following Markov chain scheme.

\vspace{2mm}

\begin{enumerate}

  \item Set initial values for all the unknown parameters in the model. We also fix the desired number of Markov iterations, with iteration index $t$.

  \item At the $t$th iteration, in accord with Figure \ref{figure3} (b), we replace the missing values of $Y$ in $B$ with $Y^{*}$ which is drawn from a Normal distribution with mean $\omega_{Y}^{(t)}+\beta_{1}^{(t)}X_{1}+\beta_{2}^{(t)}X_{2}+\beta_{3}^{(t)}X_{3}$ and standard deviation $\sigma_Y^{(t)}$, where ($X_{1}$, $X_{2}$, $X_{3}$) are observed values in dataset $B$. Similarly, we replace the missing values of $\mathbf{X}$ in $C$ with ($X_{1}^{*}$, $X_{2}^{*}$, $X_{3}^{*}$) which are drawn from a Normal distribution with mean $\omega_{1}^{(t)}+\alpha_{1}^{(t)}Z_{1},$ $\omega_{2}^{(t)}+\alpha_{2}^{(t)}Z_{2}$ and $\omega_{3}^{(t)}+\alpha_{3}^{(t)}Z_{2}+\alpha_{4}^{(t)}Z_{3}$ respectively, where ($Z_{1}$, $Z_{2}$, $Z_{3}$) are observed values in dataset $C$. We use superscript ${(t)}$ for the random parameters at the $t$th iteration, with $t=0$ representing their initial values. Because both $U$ and the random errors have zero mean, they have vanished in this step.

  \item Merge all the data, imputed and observed, into a new complete dataset (Figure \ref{figure3} (c)).

  \item Estimate model parameters based on the complete dataset obtained in Step 3 using MCMC and set  $t \leftarrow t+1$.

  \item Repeat Steps 2-4 until $t$ equals the number of iterations specified in Step 1.

  \end{enumerate}

\vspace{2mm}

\subsection*{The Priors}

Here we discuss our choice of prior distributions of the unknown parameters involved in Models (\ref{3.1})-(\ref{3.5}). Let the priors for $\bbeta = (\beta_{1}, \beta_{2}, \beta_{3})$ be independently normally distributed with mean 0 and standard deviation 10
 \begin{equation*}
  \left(
      \begin{array}{c}
        \beta_{1} \\
        \beta_{2} \\
        \beta_{3} \\
      \end{array}
    \right) \sim N\left[\left(
                     \begin{array}{c}
                        0 \\
                        0 \\
                        0 \\
                     \end{array}
                 \right),\left(
                            \begin{array}{ccc}
                                 10^2 & 0 & 0 \\
                                 0 & 10^2 & 0 \\
                                 0 & 0 & 10^2 \\
                            \end{array}
                         \right)\right],
\end{equation*}
and the IV strength parameters $\balpha = (\alpha_{1}, \alpha_{2}, \alpha_{3}, \alpha_{4})$ be independently normally distributed with mean 1 and standard deviation 0.3
\begin{equation*}
  \left(
      \begin{array}{c}
        \alpha_{1} \\
        \alpha_{2} \\
        \alpha_{3} \\
        \alpha_{4} \\
      \end{array}
    \right) \sim N\left[\left(
                     \begin{array}{c}
                        1 \\
                        1 \\
                        1 \\
                        1 \\
                     \end{array}
                 \right),\left(
                            \begin{array}{cccc}
                                 {0.3}^{2} & 0 & 0 & 0 \\
                                 0 & {0.3}^{2} & 0 & 0 \\
                                 0 & 0 & {0.3}^{2} & 0 \\
                                 0 & 0 & 0 & {0.3}^{2} \\
                            \end{array}
                         \right)\right].
\end{equation*}
$U$ has been assumed to follow a normal distribution $U \sim N(0, 0.1)$ in Model (\ref{3.1}). The priors of the standard deviation parameters $\sigma_{1}$, $\sigma_{2}$, $\sigma_{3}$ and $\sigma_{Y}$ are assumed to follow a same inverse-gamma distribution  $\sigma(\cdot) \sim$ \emph{Inv-Gamma}(3, 2).

\subsection*{Simulations}

In our simulations, according to Models (\ref{3.1})-(\ref{3.5}), we consider a total of 72 configurations including
\vspace{2mm}

\begin{itemize}

\item 6 sample overlapping rates (100\%, 80\%, 60\%, 40\%, 20\%, 0\%)

\item 2 IV strengths: $\balpha = (\alpha_{1}, \alpha_{2}, \alpha_{3}, \alpha_{4}) = \mathbf{0.5}$ and $\mathbf{0.1}$

\item 3 levels of the confounding effects of $U$ on $\mathbf{X}$: ($\delta_{1}, \delta_{2}, \delta_{3}) = \mathbf{1}, \mathbf{0.5}$ and $\mathbf{0.1}$

\item 2 levels of causal effects of $\mathbf{X}$ on $Y$: $\bbeta = (\beta_{1}, \beta_{2}, \beta_{3}) = \mathbf{0.3}$ and $\mathbf{0}.$

\end{itemize}

\vspace{2mm}

The effect of $U$ on $Y$ is set to 1 ($\delta_{4} = 1$).
In each configuration, we simulated 200 datasets, step by step, as follows.

\vspace{2mm}

\begin{enumerate}
\item Generate a dataset $\textbf{H}$ which contains observed IVs ($\mathbf{Z}$), exposures ($\mathbf{X}$) and outcome ($Y$) from 1000 independent individuals.
 \item Randomly sample $n_A$ individuals without replacement from $\textbf{H}$ and take their observations of ($\mathbf{Z}, \mathbf{X}, Y$) as dataset $A$;
  \item Randomly Sample $n_B$ individuals without replacement from $\textbf{H}-A = \{x \in \textbf{H},  x \notin A\}$ and take their observations of ($\mathbf{Z}, \mathbf{X}$) as dataset $B$;
  \item Randomly sample $n_C$ individuals without replacement from $\textbf{H}-A-B = \{x \in \textbf{H}, x\notin A \cup B\}$ and take their observations of ($\mathbf{Z}, Y$) as dataset $C$.

\end{enumerate}

\vspace{2mm}

The sample size of $B$ was set to be the same as that of $C$ (i.e., $n_B = n_C$) and the sample size of both $D_{1} = A \cup B$ and $D_{2} = A \cup C$ to 400.
The overlapping rate was defined as the percentage of the number of individuals from $A$  ($n_A$) in $D_1$ (or equivalently, in $D_2$). For example, if overlapping rate was 80\%, then $n_A = 320$ and $n_B = n_C = 80$.  Similarly, for an overlapping rate of 60\%,  $n_A = 240$ and $n_B = n_C = 160$.  When overlapping rate was 100\%, we only used dataset $A$ of sample size 400 (or equivalently, $D_1 = D_2 = A$ where $n_A = 400$). For an overlapping rate of 0\%, we only used datasets $B$ and $C$ (i.e., $D_1 = B$ with $n_B = 400$ and $D_2 = C$ with $n_C = 400$). Imputations of missing data and causal effect estimations were then performed simultaneously using MCMC in \texttt{Stan} \cite{Stan2014, Martin2008}.

\subsection*{Assessments of the performance of our method}

The performance of our method was evaluated using 4 metrics:

\vspace{2mm}

\begin{itemize}

  \item mean (posterior mean)
  \item standard deviation
  \item coverage (the proportion of the times that the 95\% credible interval contained the true value of the causal effect)
  \item power (the proportion of the times that the 95\% credible interval did not contain value zero when the true causal effect was non-zero).

\end{itemize}

\vspace{2mm}

The causal effects of $X_1$, $X_2$ and $X_3$ on the outcome $Y$ (denoted as $\hat{\bbeta} = (\hat{\beta_1}, \hat{\beta_2}, \hat{\beta_3})$) were estimated from data simulated under the alternative hypothesis that $\bbeta = \mathbf{0.3}$, and from data simulated under the null hypothesis that $\bbeta = \mathbf{0}$. Note that the metric power was only applicable under the alternative hypothesis when $\bbeta \neq \mathbf{0}$, by its definition. A high value of power indicates high sensitivity of the model to deviations of the data from the null or, equivalently, a low expected probability of false negatives.

We compared our method with the two-stage least squares (2SLS) regression \cite{Stephen2014} for one-sample MR (100\% overlap) and with the inverse-variance weighted (IVW) estimation \cite{Jack2016} for two-sample MR (0\% overlap). For overlapping samples, our method was compared with IVW, and in the latter, we only included data of non-overlapping samples to make it a two-sample MR.

\section*{Results}

Table \ref{table1} displays a summary of results obtained from the simulated data under the alternative hypothesis ($\bbeta = \mathbf{0.3}$). Each row of the table corresponds to a configuration of specified sample overlapping rate, instrument strength $\balpha$ and degree of confounding $\bdelta$. Columns are values of the four metrics of the estimated causal effects $\hat{\beta}$s obtained from our method and classic MR. Let us first focus on our method (grey columns ``Bayesian"). If overlapping rate was high (80\% or over), both the coverage and the power of the $\hat{\beta}$s were consistently high ($\geq$ 0.93). When the sample overlap was 60\% or under,  strong IVs ($\balpha = \mathbf{0.5}$) resulted in high statistical power (=1) while the impact of reduced IV strength ($\balpha = \mathbf{0.1}$) became detrimental, even more so when overlapping rate was reduced. When IVs were relatively weak ($\balpha = \mathbf{0.1}$), power increased as confounding effects $\bdelta$ decreased. Overall, where there was pleiotropy (in estimating $\beta_2$ because there were two competing paths from $Z_2$ to $Y$ via $X_2$ and $X_3$) the results were notably worse than those without pleiotropy (in estimating $\beta_1$ and $\beta_3$). As $\balpha$ increased, the averages of $\hat{\beta}$s (mean) became closer to the true value 0.3 and the variations (sd) decreased, concluding that higher IV strength reduced bias and resulted in more precise estimates. We note that despite the greater posterior uncertainty for $\beta_2$ across all the configurations (indicated in the ``sd" column), the posterior mean for this parameter remained relatively close to the true value. This is evidence that our model is reliable and ``honest about uncertainty". Compared with the classic MR, our method showed higher precision in estimated causal effect in all configurations ( ``sd" columns). When IV strength became weaker ($\balpha = \mathbf{0.1}$), our method led to higher power throughout. With highly overlapped samples (80\%), even for strong IVs, power of the estimates in classic MR (IVW) was relatively low, partly due to smaller sample size after removing data of overlapping samples.

Table \ref{table2} shows the results of the simulated data under the null ($\bbeta = \mathbf{0}$). Again, we first report results from our method (grey columns ``Bayesian"). The coverages of the three estimated causal effects $\hat{\beta}$s were high ($>$ 0.9) in all the configurations, indicating that the method does not produce a large number of false positives - an important property of MR analysis. Therefore, none of the sample overlapping rate, IV strength and confounding effect has had much negative influence on the results, although there seemed to be an increasing trend in bias and variation when IV strength decreased. In comparison with classic methods, our method provided estimates with higher precision (smaller values of ``sd"), but similar coverage.

Figures \ref{figure4} and \ref{figure5} depict, respectively, distributions of $\hat{\beta}$s for each of the IV-strength and overlapping rate configurations when the confounding effects $\bdelta = \mathbf{1}$, for $\bbeta = \mathbf{0.3}$ and $\bbeta = \mathbf{0}$. Each box represents distribution of the estimated causal effect $\beta$ based on 200 simulated datasets. Higher overlapping rate led to a gain in precision from our Bayesian method, which can be observed in all panels across the figures as the blue boxes became narrower in height as the sample overlap increased and reached minimum in the one-sample scenario (overlapping rate: 1). It is also shown that estimates were more precise when instruments were stronger (top panels vs bottom panels). Blue boxes were narrower than red boxes in height in all the panels, indicating that our method consistently outperformed classic MR in precision. Interestingly, in classic methods, variations of the estimates in one sample were smaller than those in two- and overlapping- samples, which is probably because no data were excluded in the one-sample 2SLS MR. For two- and overlapping- samples, we adopted two-sample IVW which used summary statistics and required data removal of overlapping samples. Consequently, the sample size in IVW was reduced to some extent for overlapping-samples, and the higher the overlap rate, the lower the precision.

Power of $\hat{\beta}$ for each IV-strength and overlapping rate configurations when the confounding effects $\bdelta = \mathbf{1}$ under the alternative $\bbeta = \mathbf{0.3}$ is presented in Figure \ref{figure6}. In Bayesian method, power was always equal to 1 for strong IVs (top panels) and gradually became lower as overlapping rate decreased for weak IVs (bottom panels). Compared with our method, estimates in classic MR had lower power, especially when IVs were weak. In classic MR, similar to the trend of precision in Figure \ref{figure4}, power was the highest for one sample and lowest for overlapping samples with highest overlap 80\%. Figure \ref{figure7} displays the coverage of  $\hat{\beta}$ for the same configurations and confounding effects as in Figure \ref{figure6} under the alternative $\bbeta = \mathbf{0.3}$. Overall, coverage was high ($>$ 0.85) in both of the MR methods. Bayesian method performed better for high overlapping rate while classic MR was better for low overlapping rate.
Under the alternative hypothesis, however, it would be sensible to focus on power rather than coverage.  When $\bbeta = \mathbf{0}$, it is seen from Figure \ref{figure8} that the coverage of $\hat{\beta}$ was high  ($>$ 0.91) in both Bayesian and classic MR. Our method consistently outperformed classic MR in power but not in coverage. This is partly because our method provides very precise estimates (narrow 95\% credible intervals). Thus, even for a slightly biased estimate, it is likely the credible interval does not include the true value, which will affect its performance in coverage.

\section*{Discussion}
In this paper we have proposed a Bayesian method that effectively enables a one-sample MR whether there are two overlapping samples or disjoint samples. It is noteworthy that our method has the best performance in one-sample setting, which is unsurprising because 1) our method has been developed on the basis of a Bayesian approach tailored for one-sample MR \cite{Carlo2018}; 2) the percentage of imputed data becomes lower as overlapping rate increases, and consequently, the uncertainty in data decreases. As discussed, our method provides the flexibility of combining data from previous studies with current ones to enhance statistical power, which is particularly advantageous because MR studies are often underpowered partly due to small sample sizes. It is also shown from our simulation results that pleiotropy (effect of $Z_2$ on $Y$ mediated by both $X_2$ and $X_3$) resulted in the worst estimated causal effect, because it involves two completing paths which led to higher uncertainty in estimation. This is, however, an issue commonly seen in MR studies \cite{Hemani2018, Verbanck2018}.

The precision of the estimates from our proposed method was higher than that from the classic MR. Our method also resulted in higher power of the estimates under the alternative hypothesis where the true causal effects were different from zero. Coverage of the estimates from both methods was similar under the null as well as the alternative hypotheses.

Our method is limited by the fact that we have focussed on a simple model with only three IVs and three exposures and a moderate number of configurations, although the configurations can never be exhaustive. In the real world, we often encounter much more complex data in which there are possibly many (weak and/or correlated) IVs and (correlated) exposures and outcomes, which triggers  a research topic of variable selections in future MR methodology \cite{Zuber2020}. Bayesian approach, however, is well known for its flexibility of building various models to address different scientific questions.

\section*{Conclusions}

We have developed a Bayesian MR that can be applied to any of the sample settings by means of treating missing data as unknown parameters which can then be imputed using MCMC. This is an important addition to the existing MR literature where some methods can only be used for one-sample and others for two-sample settings. Because of the nature of Bayesian inference, our model can be easily extended to tackle more complex MR analysis in medical research.

\section*{Abbreviations}

\textbf{RCT}: Randomised controlled trial \\
\textbf{MR:} Mendelian randomization \\
\textbf{IV:} Instrumental variable \\
\textbf{SNP:} Single nucleotide polymorphism \\
\textbf{BMI:} Body mass index \\
\textbf{MCMC:} Markov chain Monte Carlo \\
\textbf{IVW:} Inverse-Variance Weighted \\
\textbf{2SLS:} Two-stage least squares

\section*{Declarations}



\begin{backmatter}

\section*{Ethics approval and consent to participate}
Not applicable

\section*{Consent for publication}
Not applicable

\section*{Availability of data and material}
The code of data simulations is available from the corresponding author upon request.

\section*{Competing interests}
  The authors declare that they have no competing interests.

  \section*{Funding}
 This work was funded by Manchester-CSC. The funder had no role in study design, data generation and statistical analysis, interpretation of data or preparation of the manuscript.

\section*{Author's contributions}
   HG and CB conceived and supervised the study. LZ performed simulations and statistical analysis. LZ, HG and CB interpreted statistical results and wrote the manuscript. All authors read and approved the final manuscript.

   \section*{Acknowledgements}
Not applicable


\bibliographystyle{bmc-mathphys} 
\bibliography{bmc_biblio}      




\section*{Figures}

\begin{figure}[h!]
   \caption{\csentence{Schematic representation of the three assumptions required in Mendelian randomization.}
      1) Instrumental variable $Z$ is associated with the exposure $X$; 2) $Z$ is independent of the confounder $U$; 3) $Z$ is independent of the outcome $Y$, conditioning on $X$ and $U$.}
  \includegraphics[scale=0.5]{assumptions.pdf}\\
  \label{figure1}
\end{figure}

\begin{figure}[h!]
   \caption{\csentence{Diagram of our model and parameter settings.}
      There are three instrumental variables ($Z_{1}$, $Z_{2}$, $Z_{3}$) and three exposures ($X_{1}$, $X_{2}$, $X_{3}$).  $Z_{1}$ is associated with $X_{1}$ only. $Z_{3}$ is associated with $X_{3}$ only. $Z_{2}$ is associated with both $X_{2}$ and $X_{3}$. For simplicity, we assume there is no direct associations among the exposures and the instrumental variables are mutually independent.}
  \includegraphics[scale=0.45]{fullmodel.pdf}\\
  \label{figure2}
\end{figure}

\begin{figure}[h!]
  \caption{\csentence{Flowchart of imputations and merge of datasets $A, B$ and $C$ into a single complete dataset.} Solid rectangles in Steps (a) and (b) denote observed data. Dashed rectangles with $Y^{*(t)}$ and $\mathbf{X}^{*(t)}$ in Step (b) represent imputed values of $Y$ and $\mathbf{X}$ respectively in the specific $t$th iteration. By merging all the observed and imputed data in $A, B, C$, we obtain a single complete dataset in Step (c).}
  \includegraphics[scale=0.5]{flowchart.pdf}\\
  \label{figure3}
\end{figure}

\begin{figure}[h!]
  \caption{\csentence{Box plots of causal effects estimated from simulated data using our Bayesian method and classic methods (2SLS when sample overlapping rate = 1 and IVW otherwise) when $\bbeta=(\beta_{1}, \beta_{2},\beta_{3})=0.3$ and $\bdelta = (\delta_{1}, \delta_{2},\delta_{3}, \delta_4)=1$.} Results with two instrument strengths (top panels: $\balpha = (\alpha_1, \alpha_2, \alpha_3) = \mathbf{0.5}$ and bottom panels: $\balpha = \mathbf{0.1}$) are displayed, from left to right, for estimated causal effect of $X_1$, $X_2$ and $X_3$ on $Y$, denoted as $\hat{\beta}_1$, $\hat{\beta}_2$ and $\hat{\beta}_3$ respectively. Each panel consists results of six different sample overlapping rates ranging from 0 (leftmost) to 1 (rightmost) and each box plot represents causal effect estimated from 200 simulated datasets.}
  \includegraphics[width=17.3cm,height=9cm]{boxplotAlternative.pdf}\\
  \label{figure4}
\end{figure}

\begin{figure}[h!]
  \caption{\csentence{Box plots of causal effects estimated from simulated data using our Bayesian method and classic methods (2SLS when sample overlapping rate = 1 and IVW otherwise) when $\bbeta=(\beta_{1}, \beta_{2},\beta_{3})=0$ and $\bdelta = (\delta_{1}, \delta_{2},\delta_{3}, \delta_4)=1$.} Results with two instrument strengths (top panels: $\balpha = (\alpha_1, \alpha_2, \alpha_3) = \mathbf{0.5}$ and bottom panels: $\balpha = \mathbf{0.1}$) are displayed, from left to right, for estimated causal effect of $X_1$, $X_2$ and $X_3$ on $Y$, denoted as $\hat{\beta}_1$, $\hat{\beta}_2$ and $\hat{\beta}_3$ respectively. Each panel consists results of six different sample overlapping rates ranging from 0 (leftmost) to 1 (rightmost) and each box plot represents causal effect estimated from 200 simulated datasets.}
  \includegraphics[width=17.3cm,height=9cm]{boxplotNull.pdf}\\
  \label{figure5}
\end{figure}

\begin{figure}[h!]
  \caption{\csentence{Power from simulated data using our Bayesian method and classic methods (2SLS when sample overlapping rate = 1 and IVW otherwise) when $\bbeta=(\beta_{1}, \beta_{2},\beta_{3})=0.3$ and $\bdelta = (\delta_{1}, \delta_{2},\delta_{3}, \delta_4)=1$.} Results with two instrument strengths (top panels: $\balpha = (\alpha_1, \alpha_2, \alpha_3) = \mathbf{0.5}$ and bottom panels: $\balpha = \mathbf{0.1}$) are displayed, from left to right, for estimated causal effect of $X_1$, $X_2$ and $X_3$ on $Y$, denoted as $\hat{\beta}_1$, $\hat{\beta}_2$ and $\hat{\beta}_3$ respectively. Each panel consists results of six different sample overlapping rates ranging from 0 (leftmost) to 1 (rightmost) and each box plot represents causal effect estimated from 200 simulated datasets.}
  \includegraphics[width=17.3cm,height=9cm]{powerAlternative.pdf}\\
  \label{figure6}
\end{figure}

\begin{figure}[h!]
  \caption{\csentence{Coverage from simulated data using our Bayesian method and classic methods (2SLS when sample overlapping rate = 1 and IVW otherwise) when $\bbeta=(\beta_{1}, \beta_{2},\beta_{3})=0.3$ and $\bdelta = (\delta_{1}, \delta_{2},\delta_{3}, \delta_4)=1$.} Results with two instrument strengths (top panels: $\balpha = (\alpha_1, \alpha_2, \alpha_3) = \mathbf{0.5}$ and bottom panels: $\balpha = \mathbf{0.1}$) are displayed, from left to right, for estimated causal effect of $X_1$, $X_2$ and $X_3$ on $Y$, denoted as $\hat{\beta}_1$, $\hat{\beta}_2$ and $\hat{\beta}_3$ respectively. Each panel consists results of six different sample overlapping rates ranging from 0 (leftmost) to 1 (rightmost) and each box plot represents causal effect estimated from 200 simulated datasets.}
  \includegraphics[width=17.3cm,height=9cm]{coverageAlternative.pdf}\\
  \label{figure7}
\end{figure}

\begin{figure}[h!]
  \caption{\csentence{Coverage from simulated data using our Bayesian method and classic methods (2SLS when sample overlapping rate = 1 and IVW otherwise) when $\bbeta=(\beta_{1}, \beta_{2},\beta_{3})=0$ and $\bdelta = (\delta_{1}, \delta_{2},\delta_{3}, \delta_4)=1$.} Results with two instrument strengths (top panels: $\balpha = (\alpha_1, \alpha_2, \alpha_3) = \mathbf{0.5}$ and bottom panels: $\balpha = \mathbf{0.1}$) are displayed, from left to right, for estimated causal effect of $X_1$, $X_2$ and $X_3$ on $Y$, denoted as $\hat{\beta}_1$, $\hat{\beta}_2$ and $\hat{\beta}_3$ respectively. Each panel consists results of six different sample overlapping rates ranging from 0 (leftmost) to 1 (rightmost) and each box plot represents causal effect estimated from 200 simulated datasets.}
  \includegraphics[width=17.3cm,height=9cm]{coverageNull.pdf}\\
  \label{figure8}
\end{figure}


\section*{Tables}

\begin{sidewaystable}[h!]
\tiny
  \centering
  \caption{Causal effects estimated from simulated data using our Bayesian method and classic methods (2SLS when sample overlap = 100\% and IVW otherwise) when $\bbeta=(\beta_{1}, \beta_{2},\beta_{3})=\mathbf{0.3}$. Mean, standard deviation (sd), coverage and power are displayed for the estimated causal effects of the exposures $X_1$, $X_2$ and $X_3$ on the outcome $Y$, denoted as $\hat{\beta}_1$, $\hat{\beta}_2$ and $\hat{\beta}_3$ respectively. There are 36 configurations containing six sample overlapping rates (100\%, 80\%, 60\%, 40\%, 20\%, 0\%), two levels of IV strengths ($\balpha = (\alpha_1, \alpha_2, \alpha_3) = \mathbf{0.5}$ and $\mathbf{0.1}$) and three levels of effects of the confounder $U$ on the exposures  ($\bdelta = (\delta_1, \delta_2, \delta_3) = \mathbf{1}, \mathbf{0.5}$ and $\mathbf{0.1}$). The effect of $U$ on $Y$ ($\delta_4$) is set to 1.}
\begin{tabular}{|p{0.3cm}<{\centering}p{0.00005cm}<{\centering}p{0.2cm}<{\centering}|p{0.3cm}<{\centering}ccc|p{0.3cm}<{\centering}ccc|p{0.00005cm}|p{0.3cm}<{\centering}ccc|p{0.3cm}<{\centering}ccc|p{0.00005cm}|p{0.3cm}<{\centering}ccc|p{0.3cm}<{\centering}ccc|}
    \hline
    \multicolumn{3}{|c|}{} & \multicolumn{8}{c|}{$\hat{\beta}_{1}$}                           &       & \multicolumn{8}{c|}{$\hat{\beta}_{2}$}                           &       & \multicolumn{8}{c|}{$\hat{\beta}_{3}$} \\
\cline{1-11}\cline{13-20}\cline{22-29}    \multirow{2}[2]{*}{\textit{\textbf{overlap}}} & \multirow{2}[2]{*}{\textit{$\balpha$}} & \multirow{2}[2]{*}{\textit{$\bdelta$}} & \multicolumn{4}{c|}{\cellcolor[rgb]{ .851,  .851,  .851}\textbf{Bayesian}} & \multicolumn{4}{c|}{\textbf{Classic}} &       & \multicolumn{4}{c|}{\cellcolor[rgb]{ .851,  .851,  .851}\textbf{Bayesian}} & \multicolumn{4}{c|}{\textbf{Classic}} &       & \multicolumn{4}{c|}{\cellcolor[rgb]{ .851,  .851,  .851}\textbf{Bayesian}} & \multicolumn{4}{c|}{\textbf{Classic}} \\
\cline{4-11}\cline{13-20}\cline{22-29}          &       &       & \cellcolor[rgb]{ .851,  .851,  .851}\textbf{mean} & \cellcolor[rgb]{ .851,  .851,  .851}\textbf{sd} & \cellcolor[rgb]{ .851,  .851,  .851}\textbf{coverage} & \cellcolor[rgb]{ .851,  .851,  .851}\textbf{power} & \textbf{mean} & \textbf{sd} & \textbf{coverage} & \textbf{power} &       & \cellcolor[rgb]{ .851,  .851,  .851}\textbf{mean} & \cellcolor[rgb]{ .851,  .851,  .851}\textbf{sd} & \cellcolor[rgb]{ .851,  .851,  .851}\textbf{coverage} & \cellcolor[rgb]{ .851,  .851,  .851}\textbf{power} & \textbf{mean} & \textbf{sd} & \textbf{coverage} & \textbf{power} &       & \cellcolor[rgb]{ .851,  .851,  .851}\textbf{mean} & \cellcolor[rgb]{ .851,  .851,  .851}\textbf{sd} & \cellcolor[rgb]{ .851,  .851,  .851}\textbf{coverage} & \cellcolor[rgb]{ .851,  .851,  .851}\textbf{power} & \textbf{mean} & \textbf{sd} & \textbf{coverage} & \textbf{power} \\
\hline
    \end{tabular}%
  \label{table1}%
\end{sidewaystable}

\begin{sidewaystable}[h!]
\tiny
  \centering
  \caption{Causal effects estimated from simulated data using our Bayesian method and classic methods (2SLS when sample overlap = 100\% and IVW otherwise) when $\bbeta=(\beta_{1}, \beta_{2},\beta_{3})=\mathbf{0}$. Mean, standard deviation (sd) and coverage are displayed for the estimated causal effects of the exposures $X_1$, $X_2$ and $X_3$ on the outcome $Y$, denoted as $\hat{\beta}_1$, $\hat{\beta}_2$ and $\hat{\beta}_3$ respectively. There are 36 configurations containing six sample overlapping rates (100\%, 80\%, 60\%, 40\%, 20\%, 0\%), two levels of IV strengths ($\balpha = (\alpha_1, \alpha_2, \alpha_3) = \mathbf{0.5}$ and $\mathbf{0.1}$) and three levels of effects of the confounder $U$ on the exposures  ($\bdelta = (\delta_1, \delta_2, \delta_3) = \mathbf{1}, \mathbf{0.5}$ and $\mathbf{0.1}$). The effect of $U$ on $Y$ ($\delta_4$) is set to 1.}
\begin{tabular}{|ccc|ccc|ccc|c|ccc|ccc|c|ccc|ccc|}
    \hline
    \multicolumn{3}{|c|}{} & \multicolumn{6}{c|}{$\hat{\beta}_{1}$}           &       & \multicolumn{6}{c|}{$\hat{\beta}_{2}$}           &       & \multicolumn{6}{c|}{$\hat{\beta}_{3}$} \\
\cline{1-9}\cline{11-16}\cline{18-23}    \multirow{2}[2]{*}{\textit{\textbf{overlap}}} & \multirow{2}[2]{*}{\textit{$\balpha$}} & \multirow{2}[2]{*}{\textit{$\bdelta$}} & \multicolumn{3}{c|}{\cellcolor[rgb]{ .851,  .851,  .851}\textbf{Bayeisan}} & \multicolumn{3}{c|}{\textbf{Classic}} &       & \multicolumn{3}{c|}{\cellcolor[rgb]{ .851,  .851,  .851}\textbf{Bayeisan}} & \multicolumn{3}{c|}{\textbf{Classic}} &       & \multicolumn{3}{c|}{\cellcolor[rgb]{ .851,  .851,  .851}\textbf{Bayesian}} & \multicolumn{3}{c|}{\textbf{Classic}} \\
\cline{4-9}\cline{11-16}\cline{18-23}          &       &       & \cellcolor[rgb]{ .851,  .851,  .851}\textbf{mean} & \cellcolor[rgb]{ .851,  .851,  .851}\textbf{sd} & \cellcolor[rgb]{ .851,  .851,  .851}\textbf{coverage} & \textbf{mean} & \textbf{sd} & \textbf{coverage} &       & \cellcolor[rgb]{ .851,  .851,  .851}\textbf{mean} & \cellcolor[rgb]{ .851,  .851,  .851}\textbf{sd} & \cellcolor[rgb]{ .851,  .851,  .851}\textbf{coverage} & \textbf{mean} & \textbf{sd} & \textbf{coverage} &       & \cellcolor[rgb]{ .851,  .851,  .851}\textbf{mean} & \cellcolor[rgb]{ .851,  .851,  .851}\textbf{sd} & \cellcolor[rgb]{ .851,  .851,  .851}\textbf{coverage} & \textbf{mean} & \textbf{sd} & \textbf{coverage} \\
\hline
    \end{tabular}%
  \label{table2}%
\end{sidewaystable}

\end{backmatter}
\end{document}